\documentclass[aps,pra,epsfig]{revtex4}

\usepackage[dvips]{graphics}
\usepackage{graphicx}
\usepackage{amsfonts}
\usepackage{amssymb}
\usepackage{amsmath}

\begin{document}

\title{One-Dimensional Bose Gases with $N$-Body Attractive Interactions}
\author{E. Fersino$^{a,b}$, G. Mussardo$^{a,b,c}$, and A. Trombettoni$^{a,b}$} 
\affiliation{$^a$ International School for Advanced Studies, Trieste\\
$^b$ Istituto Nazionale di Fisica Nucleare, Sezione di Trieste \\
$^c$ Abdus Salam International Centre of Theoretical Physics, Trieste}

\begin{abstract}
We study the ground state properties of a one-dimensional Bose gas with $N$-body attractive 
contact interactions. 
By using the explicit form of the bright soliton solution of a 
generalized nonlinear Schr\"odinger equation, 
we compute the chemical potential and the ground state energy. For $N=3$, a localized soliton wave-function exists only for a critical value of the interaction strength: in this case the ground state has an infinite degeneracy that can be parameterized by the chemical potential.  
The stabilization of the bright soliton solution by an external harmonic trap 
is also discussed, and a comparison with the effect of $N$-body 
attractive contact interactions in higher dimensions is presented.
\end{abstract}
\maketitle

\section{Introduction}
\noindent
One-dimensional quantum systems have always attracted a lot of theoretical interest: their exact solutions give, in fact, useful insights on the role of the interactions and other non-perturbative features 
\cite{mattis93,korepin93,ablowitz04,giamarchi04}. In this respect, the growing ability to realize and manipulate one-dimensional Bose gases 
\cite{gorlitz01,strecker02,khaykovich02,paredes04,kinoshita04,esteve06,kinoshita06,schumm07,vandruten07} 
has provided an highly controllable experimental counterpart to these theoretical achievements.

Quasi-one-dimensional Bose gases are obtained by using a cigar-shaped external trapping potential, elongated in a direction, 
with the other degrees of freedom frozen due to the presence of a tight transverse confinement. 
In the experiments, several variants of the interacting Bose gas in one dimension 
can be implemented: an optical lattice can be added to detect the Mott-superfluid transition in one dimension \cite{kohl05}, 
the effective one-dimensional interaction can be tuned \cite{olshanii98} 
to observe a Tonks-Girardeau gas of ultracold atoms \cite{paredes04,kinoshita04}, or  
the effect of the temperature can be studied 
\cite{vandruten07}. 
Important tools that permit to further control the properties of low-dimensional Bose systems are the tuning of an 
external magnetic field near a Feshbach resonance \cite{book1,book2}, 
and, in perspective, the implementation of the recently proposed schemes to engineer 
effective three-body interactions \cite{cooper04,buchler07,paredes07}.

The technique of Feshbach resonances permits to change the sign of the scattering length: 
by switching from repulsion to attraction, i.e. from positive to negative scattering length, 
the homogeneous $1$-D Gross-Pitaevskii equation (GPE)  admits a solution corresponding to a localized wave-function, the so-called {\em bright soliton}  \cite{ablowitz04,book1,book2}.
Bright matter-wave solitons were created both in Bose-Einstein condensates of $^7$Li \cite{strecker02,khaykovich02} 
and $^{85}$Rb atoms \cite{cornish06}. 
Various localized states has been also  produced in quasi-one-dimensional geometries 
(for reviews see the book \cite{panos07}).  

With attractive two-body interactions, a crucial role is played both by the dimension of the system and the trapping potential. In three dimensions, for instance,  homogeneous attractive bosons are unstable against the collapse, but the presence of an external harmonic trap can stabilize them: the critical value of the interaction coupling that gives rise to the collapse can be obtained from the GPE \cite{book1,book2}, 
and the critical particle number is given by $\sim {\cal N}_T \mid a\mid/a_{osc}$ where ${\cal N}_T$ is the total number of particles, 
$a<0$ is the scattering length and $a_{osc}$ is the harmonic oscillator length \cite{ruprecht95,fetter95,baym96,salasnich98}.
In the one-dimensional case, the bright soliton solution is the ground state of the homogeneous GPE with negative scattering length. 
Furthermore, the GPE ground state energy is in agreement, in the thermodynamic limit, 
with the ground state energy obtained by Bethe ansatz for the attractive one-dimensional Bose gas \cite{calogero75} (see more in Section II). 

In this paper, motivated by the recent papers \cite{buchler07,paredes07} in which different schemes have been proposed to realize effective tunable 
three-body interactions, we consider an attractive three-body contact potential and, more generally, a $N$-body contact interaction. 
We consider the limit of large number of particles, ${\cal N}_T >> 1$,  with the constraint $c {\cal N}_T^{(N-1)} = const$ ($c$ being the strength of the $N$-body interactions) so that the energy per particle is finite. 
Since no Bethe solution is available in the general case of $N$-body interaction, we employ 
an Hartree approximation to study the problem in the limit mentioned above . This means that the ground state energy 
is estimated by using the bright soliton solution of a generalized mean-field GPE equation. 
As we will show, the $N=3$ is a special case: for this value, in fact, a localized soliton wavefunction exists only for a critical value of the interaction strength and has an infinite degeneracy. The stabilization of this bound state can be cured by putting the system in an external harmonic trap. 
The variational approach, that we will also employ, reveals the tendency of the higher body interactions to become more unstable in higher dimensions. It is worth stressing that the case we are considering does not consist of a $N$-body interaction added to the $2$-body interaction of the Bose gas: 
we are interested, in fact, to the effect of the $N$-body in its own, since the coefficient of the two-body interaction 
can be tuned to be zero \cite{buchler07}.

The plan of the paper is the following: in Section II we introduce the Hamiltonian corresponding to $N$-body contact attractive interactions and 
we write the (mean-field) generalized GPE. The familiar case $N=2$ is briefly recalled. 
In Section III the bright soliton solution for the homogeneous limit is obtained by using a mechanical analogy with a fictitious particle moving 
in a potential, and its properties are investigated. 
The comparison with the numerical results confirms that for $N \leq 3$ this is the ground state of the generalized GPE, as expected. 
The ground state energy by varying $N$ is also determined. 
In Section IV we consider the effect of an harmonic trap: using a variational ansatz for the ground state 
we determine the critical value of the interaction needed to stabilize the bound state. In Section V there are our conclusions.

\section{$N$-body attractive contact interactions}
\noindent
The general quantum Hamiltonian for an homogeneous one-dimensional Bose gas with $N$-body interactions $V(x_1,\cdots,x_N)$ is
\begin{equation}
\hat{H}=\int dx \hat{\Psi}^{\dag} (x) \left( -\frac{\hbar^2}{2m} \frac{\partial^2}{\partial x^2} 
\right) \hat{\Psi}(x) + \frac{1}{N!} 
\int dx_1 \cdots dx_N  \hat{\Psi}^{\dag} (x_1) \cdots \hat{\Psi}^{\dag} (x_N) 
V(x_1,\cdots,x_N)  \hat{\Psi} (x_N) \cdots \hat{\Psi} (x_1)\,\,\,,
\label{Ham-gen}
\end{equation}
where $\hat{\Psi}(x)$ is the bosonic field operator. The Lieb-Liniger Hamiltonian for the interacting one-dimensional Bose gas \cite{lieb63} has the kinetic term plus a density-density term involving pairs of particles interacting via a contact two-body potential; this corresponds to $N=2$ and $V(x_1,x_2)=V_0 \, \delta(x_1-x_2)$: $V_0$ positive (negative) corresponds to repulsion (attraction) between the bosons. The low-energy properties of the Lieb-Liniger model can be studied by the Luttinger liquid effective description \cite{haldane81} obtained by bosonization \cite{gogolin98} (a general discussion of the correlation functions is presented in \cite{cazalilla04}).

For $N$-body attractive contact interactions we set 
$V(x_1,\cdots,x_N)=-c \prod_{i=1}^{N-1} \delta(x_i-x_{i+1})$ ($c >0$): the Hamiltonian (\ref{Ham-gen}) reads then
\begin{equation}
\hat{H}=\int dx \hat{\Psi}^{\dag} (x) \left( -\frac{\hbar^2}{2m} \frac{\partial^2}{\partial x^2} 
\right) \hat{\Psi}(x) - \frac{c}{N !} 
\int dx  \hat{\Psi}^{\dag} (x) \cdots \hat{\Psi}^{\dag} (x) 
\hat{\Psi} (x) \cdots \hat{\Psi} (x) \,\,\,. 
\label{Ham-contact}
\end{equation}
In the Heisenberg representation, the equation of motion for the field operator is given by
\begin{equation}
i \hbar \frac{\partial \hat{\Psi}}{\partial t}= \left[\hat{\Psi},\hat{H} \right]=
-\frac{\hbar^2}{2m} \frac{\partial^2}{\partial x^2} \hat{\Psi} - 
c \left( \hat{\Psi}^{\dag} \right)^{N-1} \left( \hat{\Psi} \right)^{N-1} \hat{\Psi}\,\,\,. 
\label{dyn}
\end{equation}
For $N=2$, the corresponding Lieb-Liniger model is integrable and the ground state energy $E$ can be determined by Bethe ansatz \cite{mcguire64}: the final result is given by 
\begin{equation}
\frac{E}{{\cal N}_T}=-\frac{mc^2 \left( {\cal N}_T^2-1\right)}{24\hbar^2}\,\,\,,
\label{GS-Bethe}
\end{equation}
where ${\cal N}_T$ is the total number of particles. For large ${\cal N}_T$, from (\ref{GS-Bethe}) it follows that one has to keep the product $c \,{\cal N}_T = const$ in order to have a finite ground state energy per particle. Using the integrability of the $N=2$ model, the correlation functions of the attractive one-dimensional Bose gas at zero temperature were recently calculated in 
\cite{calabrese07}. 

For the three-body problem ($N=3$), no Bethe ansatz solution is available, except for a more complicate double-$\delta$ function Bose gas which can be mapped in a one-dimensional anyon gas \cite{kundu99}. Hence, to estimate the ground state energy $E$ we propose here to employ a mean-field (Hartree) approach: in this approach, the ground state energy is given in terms of the ground state energy of a generalized GPE. The same procedure will be employed for other values of $N$.  

Before we start the discussion of the general $N$-body case, let's briefly remind how this task can be successfully done for $N=2$ \cite{calogero75}. First of all, in the mean-field approximation the ground-state wavefunction is written as 
\begin{equation}
\psi_{GS}(x_1,\cdots,x_{{\cal N}_T}) \propto \prod_{i=1}^{{\cal N}_T} \psi_0(x_i)
\,\,\,, 
\end{equation}
where the function $\psi_0(x)$ is the ground state of the time-independent homogeneous GPE, i.e. the nonlinear Schr\"odinger equation (NLSE), given by
\begin{equation}
-\frac{\hbar^2}{2m} \frac{\partial^2}{\partial x^2} \psi_0-c\mid \psi_0 \mid^2 \psi_0\,=\,\mu \psi_0 \,\,\, , 
\label{GPE}
\end{equation}
where $\mu$ is the chemical potential and the normalization is given by $\int dx \mid \psi_0 \mid^2={\cal N}_T$. The energy is expressed as 
\begin{equation}
E_{GP}\,=\,\int dx \, \psi_0^*(x) \left[ 
-\frac{\hbar^2}{2m} \frac{\partial^2}{\partial x^2} - 
\frac{c}{2} \mid \psi_0(x) \mid^2 \right] \psi_0(x)\,\,\,.
\label{energy-GPE}
\end{equation}
The static bright soliton solution of (\ref{GPE}) is given by 
\begin{equation}
\psi_0(x)\,=\,\sqrt{{\cal N}_T} \frac{{\cal N}}{\cosh{\left( kx \right)}}
\,\,\,,
\label{familiarsoliton}
\end{equation}
with $k=mc{\cal N}_T/2\hbar^2$ and ${\cal N}=(1/2)\sqrt{mc{\cal N}_T/\hbar^2}$.  Substituting this expression in (\ref{energy-GPE}) one gets
\begin{equation}
\frac{E_{GP}}{{\cal N}_T}\,=\, -\frac{mc^2 {\cal N}_T^2}{24\hbar^2}\,\,\,,
\label{GS-GPE}
\end{equation}
i.e., the exact result (\ref{GS-Bethe}) apart terms $\propto 1/{\cal N}_T^2$. A comment is in order: in the homogeneous one-dimensional interacting case there is, 
strictly speaking, no condensate. However the condition $c \, {\cal N}_T = const$ implies that, for large ${\cal N}_T$, the coupling constant should scale to zero, $c \to 0$: hence, we are in a weak coupling regime where the mean-field GPE is expected to give reasonable results. In a similar way, for $c<0$ (repulsive interaction) the comparison between the exact and the GPE ground state energy shows that the latter gives the correct behaviour for $c \to 0$ while the Bogoliubov approximation gives the exact first-order corrections for small $\mid c \mid$ \cite{lieb63}.

Based on the analysis above, for general $N$ and in the limit $c \to 0$ we expect  
that a reasonable description of both the ground state properties and the low-energy dynamics is given by the mean-field generalized homogeneous GPE
\begin{equation}
i \hbar \frac{\partial \psi(x,t)}{\partial t}\,=\, \left( -\frac{\hbar^2}{2m} \frac{\partial^2}{\partial x^2} 
-c \vert \psi(x,t) \vert^{\alpha} \right) \psi(x,t)\,\,\,,
\label{GNLS}
\end{equation}
where the nonlinearity degree $\alpha$ of the generalized nonlinear 
Schr\"odinger (GNLSE) equation is related to $N$ by
\begin{equation}
N \equiv \frac{\alpha}{2} + 1\,\,\,.
\label{relation}
\end{equation}
The mean-field ground state is given by the time-independent GNLSE equation
\begin{equation}
\left( -\frac{\hbar^2}{2m} \frac{\partial^2}{\partial x^2} 
-c \vert \psi_0(x) \vert^{\alpha} \right) \psi_0(x) \,=\, \mu \psi_0(x)\,\,\,,
\label{mu}
\end{equation}
where, as before, $\mu$ is chemical potential and $\psi_0$ is normalized to the total number of particles ${\cal N}_T$, i.e. $\int dx \vert \psi_0(x) \vert^2={\cal N}_T$. 

With $N \ge 2$ integer, $\alpha$ is an even integer; however, in eqn. (\ref{mu}) 
$\alpha$ can take any real positive value and, in the following, we will consider this general case. In this respect let's comment that the axial dynamics of a Bose-Einstein condensate induced by an external potential with cylindrical symmetry in the transverse directions can be studied by introducing an effective one-dimensional GPE equation with $\alpha=1$ \cite{salasnich02} and that for Bose-Einstein condensates in one-dimensional optical lattices the effective equation has a value of $\alpha$ that depends on the details of the trapping potentials and it is, in general, a non-integer value \cite{smerzi03}.

\section{Ground state of the generalized nonlinear Schr\"odinger equation}
\noindent
In the following we will study the attractive $N$-body problem in the thermodynamic limit, defined by 
${\cal N}_T \rightarrow \infty$, with the product  
$G = c {\cal N}_T^{\alpha/2} $ kept fixed. This will ensure the energy per particle of the GNLSE bright soliton to be finite. In dimensionless units, rescaling the wave function $\psi_0 \rightarrow \sqrt{{\cal N}_T} \psi_0$, eqn. (\ref{mu}) reads
\begin{equation}
\left( -\frac{1}{2} \frac{\partial^2}{\partial x^2} 
-g \vert \psi_0(x) \vert^{\alpha} \right) \psi_0(x) \,=\, \tilde\mu \psi_0(x) \,\,\,,
\label{mu-ad}
\end{equation}
where $g$ and $\tilde\mu$ are the dimensionless versions of $G$ and $\mu$ respectively. For the ground state of this equation we look for a real solution, with the normalization condition 
\begin{equation}
\int_{-\infty}^{\infty} \psi_0^2(x) \,dx\, =\,1 \,\,\,.
\label{important}
\end{equation} 
Obviously, once a static solution $\psi_0(x)$ of eqn. (\ref{mu-ad}) has been found, 
the corresponding soliton wave solution with velocity $v$ is given by 
\begin{equation}
\psi_0(x,t)\,=\,\psi_0(x-vt) \, e^{-i (\tilde\mu t-v x+v^2t/2)}\,\,\,.
\label{sol-wav}
\end{equation}
The solution of eqn. (\ref{mu-ad}) can be found by using a mechanical analogy. 
In fact, interpreting $x$ as the time variable and $\psi_0(x)$ as the coordinate of a fictitious particle, 
eqn. (\ref{mu-ad}) formally corresponds to the Newton's equation of motion of this particle (of mass $M = 1/2$), subjected to the force
\begin{equation}
F\,=\, - \tilde\mu \psi_0 - g \psi_0^{\alpha+1} \,\,\,.
\end{equation}
This force can be derived by the potential 
\begin{equation}
V(\psi_0)\,=\,\frac{\tilde\mu}{2} \psi_0^2 +\frac{g}{\alpha+2} 
\psi_0^{\alpha+2} \,\,\,.
\label{potential}
\end{equation}
As any motion of a particle in a potential, this is accompanied by the integral of motion that corresponds to its mechanical energy 
\begin{equation}
{\cal H} \,=\,\frac{M}{2} \left(\frac{d\psi_0}{dx}\right)^2 + 
V(\psi_0) \,=\,{\rm const} \,\,\,.
\label{energymech}
\end{equation}

\begin{figure}[t]
\begin{center}
\includegraphics[width=4.cm,height=4.cm,angle=270,clip]{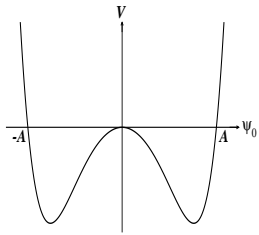} 
\caption{Typical shape of the potential $V(\psi_0)$ 
for negative values of $\tilde\mu$. 
$\pm A$ are the inversion points of the motion.}
\label{potentialfigure}
\end{center}
\end{figure} 

\noindent
Following this mechanical analogy, it is easy to see that a non-trivial motion can take place only if $\tilde\mu < 0$, where the typical shape of the potential is similar to the one drawn in Fig. \ref{potentialfigure}. 

Notice that a solution is always given by the equilibrium configuration of the potential (\ref{potential}), i.e. 
\begin{equation}
\psi_0(x)\,=\,\left(\frac{-\tilde\mu}{g}\right)^{1/\alpha} \,\,\,.
\label{constantsolution}
\end{equation}
This solution can be normalized only on a finite volume $L$, with the dependence of the chemical potential on the volume determined by the normalization condition (\ref{important}), i.e.  $\tilde\mu=-g L^{-\alpha/2}$: given this dependence of the chemical potential, the constant solution is simply 
\begin{equation}
\psi_0 \,=\,\frac{1}{\sqrt{L}} \,\,\,.  
\end{equation}
To determine the ground state, we have to compare the GPE energy of this constant solution with the one of a localized wavefunction.
For this solution, both $\psi_0(x)$ and $\frac{d\psi_0}{dx}(x)$ should vanish when $x\rightarrow \pm \infty$. 
This condition fixes the constant value of ${\cal H}$ to be zero (notice that this value is $not$ the GPE energy). 
In this case, the fictitious particle takes off from the origin at $x = -\infty$, moving to the right 
(or, equivalently to the left, since the original equation is invariant under $\psi_0(x) \rightarrow - \psi_0(x)$), 
until it reaches the inversion point $A$ at the time $x=0$. Once the particle arrives in $A$, 
it inverts its motion and comes back to the origin with a vanishing velocity. It is clear from this analogy that $A$ will be the maximum of the bright soliton solution. 

What we said, however, is not the end of the story. 
In fact, the kind of motion we have just described occurs for any potential with the shape shown in Fig. \ref{potentialfigure}. But we are looking for that particular motion that satisfies the additional constraint (\ref{important}) and this condition can be fulfilled only for a particular shape of the potential, i.e. for a particular combination of the parameters $\tilde\mu$ and $g$: it is as if the solution is looking for its proper potential.  
     
\begin{figure}[t]
\begin{center}
\includegraphics[width=6.cm,height=8.cm,angle=270,clip]{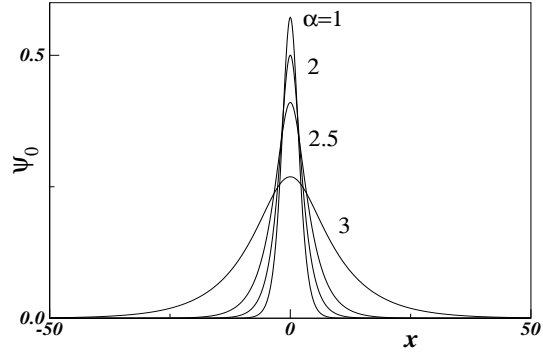} 
\caption{Ground state wavefunctions of the GNLSE (\ref{mu-ad}) for $\alpha=1,2,2.5,3$ (dimensionless units are used with $g=1$).}
\label{GSS}
\end{center}
\end{figure}

In the following, it is convenient to introduce the quantities 
\begin{equation}
a^2 \equiv -\frac{\tilde\mu (\alpha+2)}{2 g} > 0 
\,\,\,\,\,\,\,
,
\,\,\,\,\,\,\,
b \equiv \sqrt{\frac{4 g}{\alpha +2}} 
\,\,\,\,\,\,\,
,
\,\,\,\,\,\,\,
\gamma \equiv \frac{2}{\alpha}
\,\,\,.
\end{equation}
Using the integral of motion ${\cal H}$, for generic values of $\tilde\mu$, $g$ and $\alpha$ (with $\tilde\mu < 0$,  $g >0$ and $\alpha >0$) 
the solution is given by a quadrature  
\begin{equation}
\int_{A}^{\psi_0(x)} \frac{dq}
{\sqrt{a^2 q^2 - q^{\alpha +2}}} \,=\,
b \, \int_0^x d \tau 
\,\,\,,
\label{quadrature}
\end{equation}
where $A = a^{\gamma}$ is the inversion point reached by the particle at the "time" $x = 0$. Using the exact expression of the integral of the left hand side 
\begin{equation}
\int_{A}^{\psi_0(x)} 
\frac{dq}{\sqrt{a^2 q^2 - q^{\alpha +2}}} \,=\,
-\frac{1}{a \alpha} \, {\rm ln} 
\left[
\frac{a+\sqrt{a^2 - \psi_0^{\alpha}(t)}}
{a-\sqrt{a^2 - \psi_0^{\alpha}(t)}}
\right] \,\,\,, 
\label{exactintegral}
\end{equation}
one gets 
\begin{equation}
\psi_0(x)\,=\,\frac{A}{\cosh^{\gamma}{\left( \frac{\alpha}{2} \sqrt{-2\tilde\mu} \,x \right)}} \,\,\,.
\label{q-result}
\end{equation}
It remains now to impose the normalization eqn. (\ref{important}) to the solution (\ref{q-result}) : this fixes the shape of the potential, i.e. the relation between $\tilde\mu$ and $g$  
\begin{equation}
\left( - \tilde\mu \right)^{\frac{4-\alpha}{2\alpha}}=g^{2/\alpha} \, \left( \frac{2}{\alpha+2} \right)^{2/\alpha} \, \frac{\alpha \Gamma(2/\alpha+1/2)}
{\sqrt{2 \pi} \Gamma(2/\alpha)} \,\,\,.
\label{relation-mu-g}
\end{equation}
When $\alpha \neq 4$, we can use this equation to express $\tilde\mu$ as a function of $g$ and, in particular, to write the normalization $A$ as 
\begin{equation}
A\,=\, \left( \sqrt{\frac{2 g}{\pi \gamma (\gamma+1)}} 
\frac{\Gamma(\gamma+1/2)}{\Gamma(\gamma)}\right)^{\frac{2}{4-\alpha}}\,\,\,.
\label{enne}
\end{equation}
In Fig. \ref{GSS} we plot the soliton solution (\ref{q-result}) for different values of $\alpha$ at $g=1$. 

\begin{figure}[t]
\begin{center}
\includegraphics[width=6.cm,height=8.cm,angle=270,clip]{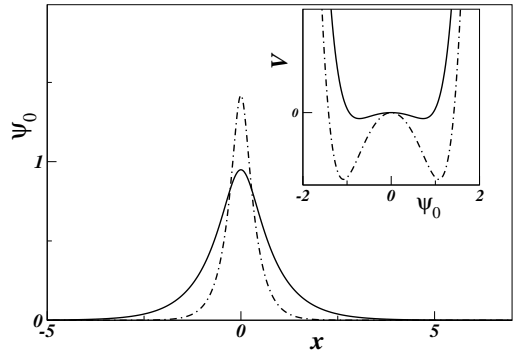} 
\caption{Wavefunction (\ref{psialpha4}) for $N=3$ and $g=g^\ast$ 
plotted for $\tilde\mu=-1$ (solid line) and $\tilde\mu=-5$ (dot-dashed line). Inset: corresponding potential 
(\ref{potential}) for $\tilde\mu=-1$ (solid line) and $\tilde\mu=-5$ (dot-dashed line).}
\label{mu-pot}
\end{center}
\end{figure}

However, when $\alpha=4$, eqn. (\ref{relation-mu-g}) leaves $\tilde\mu$ undetermined: this means that the corresponding wavefunction 
\begin{equation}
\psi_0(x) \,=\,\left( \frac{\sqrt{-3 \tilde\mu/g}}{\cosh(2\sqrt{-2\tilde\mu} \, x)} \right)^{1/2}\,\,\,
\label{psialpha4}
\end{equation}
is solution of the nonlinear Schr\"odinger equation (\ref{mu-ad}) for {\em every} $\tilde\mu$. In this case, however, only a particular value of $g$, given by 
\begin{equation}
g^\ast\,=\,\frac{3\pi^2}{8} \,\,\,,
\label{criticalc}
\end{equation}
guarantees its correct normalization (\ref{important}). Expressed in more physical terms, 
the attractive $3$-body interaction has the peculiarity that one can arbitrarily vary the chemical potential provided that the coupling constant be fine-tuned to the critical value $g^\ast$: 
increasing or decreasing (in modulus) the chemical potential simply results, in this case, 
in shrinking or enlarging the shape of the soliton. This is shown in Fig. \ref{mu-pot} where the wavefunction 
(\ref{psialpha4}) is plotted for two different values of $\tilde\mu$: in the inset we plot the corresponding potential 
(\ref{potential}), showing a larger (smaller) inversion point corresponding to the smaller (larger) width.

The fact that $\tilde\mu$ is undetermined and 
the soliton can change arbitrarily its shape does not imply that the GPE energy is undetermined: 
in fact, the explicit computation of the next subsection shows that, in this case, the energy does not depend on the value of $\tilde\mu$. 
Then, for $N=3$ and $g=g^\ast$, an infinite degeneracy parametrized 
by the chemical potential $\tilde\mu<0$ occurs. 

\begin{figure}[t]
\begin{center}
\includegraphics[width=6.cm,height=8.cm,angle=270,clip]{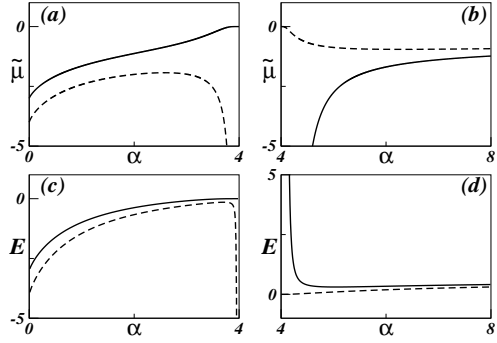} 
\caption{(a)-(c) Chemical potential and energy (in dimensionless units) of the bright soliton solution (\ref{q-result}) 
for $g=3<g^\ast$ (solid line) 
and $g=4>g^\ast$ (dashed line) for $\alpha<4$ - 
(b)-(d) chemical potential and energy for $g=3$ (solid line) and $g=4$ (dashed line) for $\alpha>4$.}
\label{mu-en}
\end{center}
\end{figure}

In Fig. \ref{mu-en}(a)-(b) we plot the chemical potential $\tilde\mu$ for two different values 
of $g$, one smaller than $g^\ast$ and the other larger: it is seen that for $g<g^\ast$ ($g>g^\ast$), then  
$\tilde\mu \to 0$ ($\tilde\mu \to - \infty$) for $\alpha \to 4^-$ 
while $\tilde\mu \to \infty$ ($\tilde\mu \to 0$) for $\alpha \to 4^+$. 
The singular nature of the $3$-body interaction can then be recovered by studying the limit $\alpha \rightarrow 4$ of the formulas (\ref{q-result}), 
(\ref{relation-mu-g}), (\ref{enne}) given above: for 
$\alpha \rightarrow 4^-$, if $g = g^\ast$ the normalization $A$ goes to 1, while if $g < 
g^\ast$, $ A \rightarrow 0$ and $\tilde\mu \rightarrow 0$ (i.e. we have a non-localized solution) 
whereas if $g > g^\ast$, both $A$ and $\tilde\mu$ diverge, i.e. the wavefunction collapses to the origin. It is worth to mention that 
a singular behavior of the nonlinear Schr\"odinger equation relative to 
the value $\alpha = 4$ has been also observed in the dynamical blowing up of the moving wave-packets of this equation: 
the interested reader is referred to the  mathematical literature for a detailed discussion of this issue \cite{merle04}. 
In the present application, 
this instability means that the local $3$-body local attractive interactions 
cannot sustain a bound state unless there is a fine tuning of the interaction. 
In the next Section we will show how an external trap can help to stabilize the bound state for a generic value of the coupling.

\begin{figure}[t]
\begin{center}
\includegraphics[width=6.cm,height=8.cm,angle=270,clip]{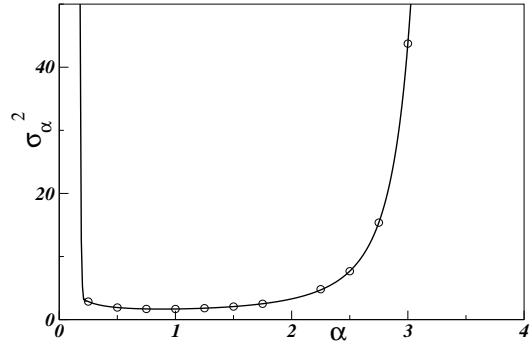} 
\caption{$\sigma_{\alpha}^2$ versus the nonlinearity degree $\alpha$. Solid line: eqn. (\ref{sigma}); open circles: $\langle x^2 \rangle$ 
from the numerical determination of the ground state of the GNLSE (\ref{mu-ad}). The value $g=1<g^\ast$ and dimensionless units are used.}
\label{DISP}
\end{center}
\end{figure}

To understand better the behaviour of the solution $\psi_0(x)$ as a function of $\alpha$, let's define the width $\sigma_\alpha$ as 
$\sigma_\alpha^2= \int dx x^2 \psi_0^2(x)$. One gets
\begin{equation}
\sigma_\alpha^2=\frac{\Gamma(\gamma+1/2)}{\pi \Gamma(\gamma)} \cdot
\frac{\gamma (\gamma+1)}{2 { A}^{\alpha}} \cdot {\cal I}_\alpha \,\,\,,
\label{sigma}
\end{equation}
where ${\cal I}_\gamma=\int  dX \, X^2 / \cosh^{2 \gamma}{(X)}$.
One finds $\sigma_2^2(g)=\pi^2/3g \approx 3.28/g$ and $\sigma_1^2(g)=(\pi^2-6)/(12g)^{1/3} \approx 1.69/g^{1/3}$. 
For large $\alpha$ one has $\sigma_\alpha^2 \to g/2$, while, of course, $\sigma_\alpha^2 \to \infty$ 
for $\alpha \to 0$ (no localized soliton without interaction). 
For $g<g^\ast$, from eqn. (\ref{sigma}) one sees that for $\alpha \to 4^-$, $\sigma_\alpha \to \infty$, 
while for $\alpha \to 4^+$, $\sigma_\alpha \to 0$. In Fig. \ref{DISP} we plot $\sigma_\alpha^2$ for $g=1<g^\ast$ from 
eqn. (\ref{sigma}) and, for completeness, also the widths for some value of $\alpha$ obtained from the numerical GNLSE.  
A divergence is observed for $\alpha \to 4^-$, corresponding to the $3$-body attraction: 
the bright soliton becomes larger and larger getting close to $\alpha=4$, while for $\alpha$ slightly larger 
than $4$ the soliton becomes extremely narrow. This means that there is a collapse of the solution (\ref{q-result}) 
going to $\alpha=4$ from large values of $\alpha$. At variance, 
for $g>g^\ast$, then for $\alpha \to 4^-$, $\sigma_\alpha \to 0$, 
while for $\alpha \to 4^+$, $\sigma_\alpha \to \infty$. 
It should be stressed that, for $\alpha>4$, although (\ref{q-result}) is a solution of the GNLSE (\ref{mu-ad}), 
it is no longer its ground state: the divergence of $\sigma_\alpha$ for $\alpha \to 4^-$ 
is signaling the disappearance of the bound state due to the $3$-body interaction.

\subsection{Ground state energy}
\begin{figure}[t]
\begin{center}
\includegraphics[width=6.cm,height=8.cm,angle=270,clip]{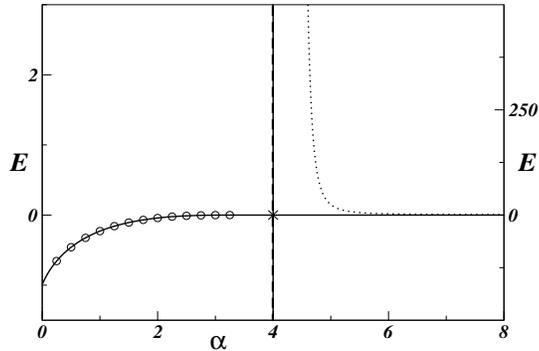} 
\caption{Ground state energy  vs. $\alpha$. 
Solid line (dashed line): eqn. (\ref{energy-GGPE}) for $\alpha<4$ ($\alpha>4$); open circles: 
energy of the numerical ground state of the GNLSE (\ref{mu-ad}). 
Dimensionless units (with ${\cal N}_T=1$) and $g=1<g^\ast$, 
as well as different scales of the energy for $\alpha<4$ (left part) 
and $\alpha>4$ (right), are used. }
\label{ENER}
\end{center}
\end{figure}

\noindent
Using the bright soliton solution (\ref{q-result}) we can now estimate the energy per particle. 
Going back to the physical dimensions of all quantities and 
normalizing now $\psi_0$ to ${\cal N}_T$, for $\alpha \neq 4$ the chemical potential is given by
\begin{equation}
\mu=-\frac{\hbar^2 \gamma^2}{2mf_\gamma^{2\alpha/(4-\alpha)}} 
\left( \frac{2mG}{\hbar^2 \gamma(\gamma+1)}\right)^{\frac{4}{4-\alpha}}\,\,\,,
\label{mu-dim}
\end{equation}
where 
$$
f_\gamma=\frac{\sqrt{\pi} \, \Gamma(\gamma)}{\Gamma(\gamma+1/2)}\,\,\,.
$$ 
The energy per particle is then obtained from the GNLSE energy functional
\begin{equation}
E_{GP}\,=\,\int dx \, \psi_0^*(x) \left[ 
-\frac{\hbar^2}{2m} \frac{\partial^2}{\partial x^2} - 
\frac{2c}{\alpha+2} \mid \psi_0(x) \mid^\alpha \right] \psi_0(x)\,\,\,.
\label{energy-GGPE}
\end{equation}
Using eqn. (\ref{q-result}) we obtain 
\begin{equation}
\frac{E_{GP}}{{\cal N}_T}\,=\,-\frac{\hbar^2}{2m} 
\left( \frac{2mG}{\hbar^2}\right)^{\frac{4}{4-\alpha}} {\cal E}(\alpha)\,\,\,,
\label{en-dim}
\end{equation}
where
\begin{equation}
{\cal E}(\alpha)\,=\,\left\{ \frac{1}{\gamma (\gamma+1) f_\gamma^2} \right\}^{\frac{4}{4-\alpha}} \cdot 
\left[ \gamma^2 f_\gamma^2 -\frac{\alpha}{\alpha+2} \gamma (\gamma+1) f_\gamma f_{\gamma+1} \right]\,\,\,.
\label{en-fun}
\end{equation}
For $N=2$, it is $\mu=-m c^2 {\cal N}_T^2/8\hbar^2$ and the previous energy (\ref{GS-GPE}) is recovered. 
From eqn. (\ref{en-dim}) it follows that in order to maintain finite the energy per particle for large ${\cal N}_T$ one has to keep $G$ fixed. 
By a numerical determination of the ground state of the GNLSE, 
we have verified that (\ref{q-result}) indeed coincides with the ground state for $\alpha < 4$ {\em both for $g<g^\ast$ and $g>g^\ast$}.
In Fig. \ref{ENER} we compare for $g=1<g^\ast$ 
the ground state energy per particle from eqn. (\ref{en-dim}) with the ground state energy obtained for some values 
of $\alpha$ obtained from the numerical GNLSE. For $g>g^\ast$ and $\alpha<4$, a similar agreement is obtained. 

For $\alpha=4$, as discussed in the previous section, the chemical potential 
is undetermined. However, a direct substitution of (\ref{psialpha4}) in 
(\ref{energy-GGPE}) reveals that $E_{GP}=0$ for $g=g^\ast$. 
Since (\ref{psialpha4}) is a solution of the GNLSE (\ref{mu-ad}) for 
arbitrary $\mu<0$, and then with arbitrary width, we conclude that 
an infinite degeneracy - parametrized 
by a negative chemical potential - occurs. 

Using the energy (\ref{en-dim}) we can also estimate the energy of the constant solution in the finite interval $[ -L/2,L/2 ]$: it is $E_{const}/{\cal N}_T=-2c \rho^{\alpha/2}/(\alpha+2)$, where 
$\rho={\cal N}_T/L$ is the density. To compare this energy with the previous result (\ref{en-dim}) for the bright soliton solution 
we have to choose how to perform the thermodynamic limit: if we choose to keep fixed the quantity $G = c\,{\cal N}_T^{\alpha/2}$, with large but finite value of ${\cal N}_T$, sending $L$ to infinite the energy of the constant solution vanishes.  This means that for $\alpha>4$ the constant solution is the ground state of the system; $\alpha=4$ and  $g=g^\ast$ is the case in which both the solutions have zero energy. 
 
\section{Effect of an harmonic trap}
\noindent
In three dimensions, the presence of an external harmonic trap (with frequency $\omega$) can help to  stabilize the attractive two-body interaction: the critical value ${\cal N}_T^{(cr)}$ of the particle number that induces the collapse, obtained from the GPE, is given by $0.57 {\cal N}_T \mid a\mid/a_{osc}$ \cite{ruprecht95}. This critical value can be estimated by a variational method 
\cite{fetter95,book1}: a gaussian trial wavefunction, with the width variational parameter, is introduced  and the corresponding energy computed \cite{baym96}. Without external trap ($\omega=0$), the energy does not have a minimum. However, with $\omega \neq 0$, two situations are possible: for a number of particle ${\cal N}_T<{\cal N}_T^{(cr)}$ a metastable minimum appears, while for ${\cal N}_T > {\cal N}_T^{(cr)}$ there is no minimum. 

In this Section we consider the corresponding problem in one dimension with a $3$-body interaction and we show that there is a critical value 
$c^{\ast}$ of the interaction, such that for $c<c^{\ast}$, the bound state is stable. To this aim, we use the variational wavefunction 
\begin{equation}
\psi_V(x) \,=\, C\,\exp(- x^2/ \sigma^2)\,\,\,,
\label{VAR}
\end{equation} 
normalized to ${\cal N}_T$. The energy to be minimized is obtained by inserting the variational wavefunction (\ref{VAR}) in the generalized GPE functional
\begin{equation}
E \,=\, \int dx \psi^*(x) \left[ -\frac{\hbar^2}{2m} \frac{\partial^2}{\partial x^2} - \frac{2 c}{\alpha +2} 
\mid \psi(x) \mid ^{\alpha}  +\frac{m}{2} \omega^2 x^2 \right]\psi(x) \,\,\,.
\label{E-functional}
\end{equation}

To better illustrate the peculiarities of the one-dimensional case, it is useful to perform the analysis also in higher dimensions. The energy in $D=1,2,3$ is given by 
\begin{equation}
\frac{E}{{\cal N}_T}\,=\, D \frac { \hbar^2}{2 m \sigma^2 } - c f_{\alpha,D}\frac {{\cal N}_T^{\frac{\alpha}{2}}}
{\sigma^{\frac{D \alpha}{2}}} +D \frac{m \omega^2 \sigma^2}{8}\,\,\,,
\label{energy}
\end{equation}
where
\begin{equation}
f_{\alpha, D} \equiv 
\frac{2}{\alpha+2} \left( \frac{\pi}{\alpha+2} \right)^{\frac{D}{2}} 
\left( \frac{2}{\pi} \right)^{D \frac{\alpha+2}{4}}\,\,\,.
\label{effe}
\end{equation}

\begin{figure}[t]
\begin{center}
\includegraphics[width=6.cm,height=8.cm,angle=270,clip]{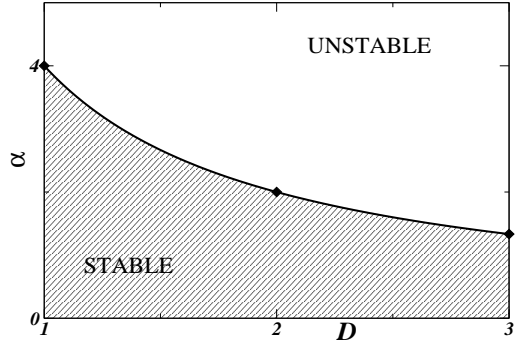} 
\caption{Stability region according to eqn. (\ref{cond-crit}): $D=1,2,3$ corresponds respectively to $\alpha=4,2,4/3$, 
i.e. $N=3,2,5/3$.}
\label{crit-fig}
\end{center}
\end{figure}

Let us consider initially the homogeneous case: for $\omega = 0$ the energy (\ref{effe}) has a minimum only when $D \,\alpha<4$.  For $D=1$, the critical value corresponds to $\alpha=4$: this is in agreement with the result of the previous Section, which is now obtained by a  variational approach. The critical condition
\begin{equation}
D \,\alpha\,=\,4\,\,\,
\label{cond-crit}
\end{equation}
is plotted in Fig.\ref{crit-fig}: this figure shows that the higher $N$-body interactions tend to be more unstable in higher dimensions.

\begin{figure}[t]
\begin{center}
\includegraphics[width=6.cm,height=8.cm,angle=270,clip]{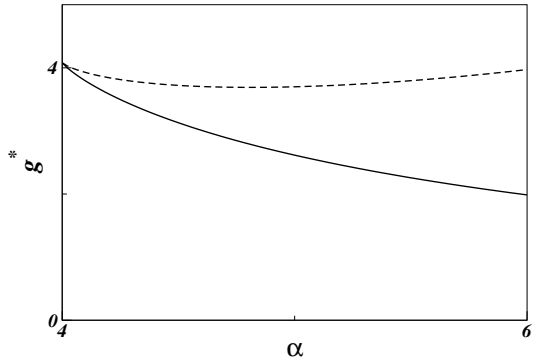} 
\caption{Critical value $g^\ast$ vs. $\alpha$ for $\alpha>4$, with $D=1$. Dimensionless units (with $g={\cal N}_T=1$) 
are used, with $\omega=1$ (dashed line) and $\omega=4$ (solid line).}
\label{plot-c-crit}
\end{center}
\end{figure}

When the harmonic trap is present ($\omega \neq 0$), there is still a minimum when the values 
$\alpha$ is pushed to the range of value $\alpha<4/D$. When $D \alpha>4$, we can identify the critical value $c^{\ast}$ as indicated in \cite{fetter95} for the $D=3$ and $\alpha=2$ case: since $E \to -\infty$ for $\sigma \to 0$ and $E \to \infty$ for $\sigma \to \infty$, the critical value is obtained by the conditions 
$\partial E / \partial \sigma = \partial^2 E / \partial \sigma^2 = 0$. In this way we arrive to the result        
\begin{equation}
c^{\ast} {\cal N}_T^\frac{\alpha}{2} \,=\, \left( \frac{ D\alpha-4}
{4 m \omega^2} \right)^\frac{D \alpha-4}{8} \left( \frac{16\hbar^2}
{m (D \alpha+4)} \right)^\frac{D \alpha+4}{8} \frac{\alpha+2}{2\alpha} 
\left( \frac{\alpha+2}{\pi} \right)^\frac{D}{2} \left( \frac{\pi}{2} \right)^{D\frac{\alpha+2}{4}} \,\,\,.
\label{g_critical}
\end{equation}
The instability curve (\ref{g_critical}) depends on $D$: with dimensionless units (and ${\cal N}_T=1$), 
in one dimension as $\alpha \to \infty $ the critical value $g^\ast$ goes to zero for $\omega \geq \pi$ and to infinity otherwise; in two dimensions the behaviour is similar except that 
$g^{\ast} \to \pi/ e$ when $\omega= \pi$; 
while in three dimensions critical value goes to infinity for  $\pi \geq \omega$ and to zero otherwise. A plot of the $g^{\ast}$ in $D=1$ for $\alpha>4$ is presented in Fig. \ref{plot-c-crit}. 

Let's now examine the critical point $D\,\alpha=4$, which in $D=1$ corresponds to the $3$-body interactions. Minimizing the energy (\ref{energy}) with respect to $\sigma$, one finds that there is the critical value 
\begin{equation}
c^{\ast}{\cal N}_T^{\alpha/2} \,=\,  \frac{D \hbar^2}{2mf_{\alpha,D}} \,\,\,.
\label{g_critical_cr}
\end{equation}
For $c < c^{\ast}$ there is a minimum and the system is stable, while for $c^{\ast}$ the energy does not have ever a minimum: hence, irrespectively of how large  $\omega$ may be, the system always collapses. For $\omega=0$ (no trap) and $c < c^\ast$ ($c>c^\ast$), the minimum value of the energy is then obtained for $\sigma=0$ ($\sigma=\infty$). Notice that the variational approach gives for $D=1$ and $\alpha = 4$, 
the critical value (in dimensionless units) 
$g^\ast = 3\sqrt{3}\pi/4\approx 4.08$, in good agreement with the analytical value 
$g^\ast = 3\pi^2/8\approx 3.70$.

\section{Conclusions}
\noindent
In this paper we have analyzed the one-dimensional Bose gases with $N$-body local attractive interactions: 
by using a mean-field approach, we found that $N=3$ (i.e., $\alpha=4$) is a critical point, and that the localized solution 
is possible only for a critical value $c^\ast$ of the interaction strength. For this critical value, 
an infinite degeneracy occurs: this degeneracy is parametrized by the chemical potential 
(i.e., eigenfunctions with the same negative $\mu$ has the same energy). For $\alpha<4$, 
the bright soliton coincides with the ground state of the Gross-Pitaevskii equation: when $c<c^\ast$ ($c>c^\ast$), then 
$\alpha \to 4^-$ gives a width going to diverge (vanish). 
We have also studied how an harmonic trap can make stable this bound state, 
pointing out that for $N=3$ there exist a critical value of the $3$-body interaction strength, and below such critical value 
the localized state occurs. 
Above this critical value, the collapse is not prevented even for very large trap frequency. 
A brief discussion of the role played by the dimension for $N$-body local attractive interactions has been also presented, 
showing that higher body interactions are more unstable in higher dimensions. 
We also mention that two-body nonlocal attractive interactions has been studied, showing different ranges 
of stability with respect to the local ones \cite{salasnich98}: we could then expect that for $3$- and $N$-body 
interactions this effect could become even more relevant.

Several proposals have recently addressed the issue of inducing and controlling three-body terms. In \cite{buchler07} it has been proposed to use cold polar molecules driven by microwave fields to obtain strong three-body interactions, controllable in a separate way from the two-body interactions, which in turn can be switched off \cite{buchler07}. Three-body interactions can be effectively induced in mixtures of bosonic particles and molecules: in \cite{cooper04} the ground state of rotating Bose gases close to a Feshbach resonance has been studied, showing that for suitable parameters they are fractional quantum Hall states, whose excitations  obey non-abelian exchange statistics. In \cite{paredes07} it was shown that a system of atoms and molecules in a one-dimensional lattice can be effectively modeled by a three-body local (i.e., contact) interaction, characterized by a strength $U$ and in the limit $U \to \infty$ (without a two-body interaction) the ground state properties were investigated by a Pfaffian-like ansatz. The strength $U$ of the three-body interaction can be made also negative by using Feshbach resonances.

One of the main reasons of interest of these proposals relies on the fact that exotic quantum phases, such as topological phases, appear to be ground states of Hamiltonian with three or more body interaction terms, an example being the fractional quantum Hall states described by the Pfaffian wavefunctions \cite{moore91}. The excitations of Pfaffian states are non-abelian anyons, on which schemes of fault-tolerant topological quantum computation are based \cite{kitaev03}. We think that, in perspective, the possibility to induce and tune effective $3$-body interactions could become an important tool to control the nonlinear dynamical properties of localized wave-packets and to induce new exotic strongly correlated phases in ultracold atoms.

\vspace{3mm}

{\em Acknowledgments}. We would like to thank B. Dubrovin and P. Calabrese for stimulating discussions. This work is partially supported 
by the ESF grant INSTANS and by the MIUR projects 
``Quantum Field Theory and Statistical Mechanics in Low Dimensions'' and ``Quantum Noise in Mesoscopic Systems''.

\end{document}